\documentstyle[aps,epsf,prl]{revtex} 

\makeatletter
\newbox\slashbox \setbox\slashbox=\hbox{$/$}
\newbox\Slashbox \setbox\Slashbox=\hbox{\large$/$}      
\def\pFMslash#1{\setbox\@tempboxa=\hbox{$#1$}
  \@tempdima=0.5\wd\slashbox \advance\@tempdima 0.5\wd\@tempboxa
  \copy\slashbox \kern-\@tempdima \box\@tempboxa}
\def\pFMSlash#1{\setbox\@tempboxa=\hbox{$#1$}
  \@tempdima=0.5\wd\Slashbox \advance\@tempdima 0.5\wd\@tempboxa
  \copy\Slashbox \kern-\@tempdima \box\@tempboxa}

\def\FMSlash{\protect\pFMSlash}
\def\miss#1{\ifmmode{/\mkern-11mu #1}\else{${/\mkern-11mu #1}$}\fi}
\makeatother

\begin{document} 

\draft 
\preprint{CINVESTAV--98--10}
\title{The Decay $Z \to \bar{\nu}\nu \gamma$ in the Standard Model}
\author{J. M. Hern\'andez} 
\address{Facultad de Ciencias F\'{\i}sico Matem\'aticas, Universidad
Aut\'onoma de Puebla, Apartado Postal 1152, Puebla, M\'exico.}
\author{M. A. P\'erez \thanks{E--mail: mperez@fis.cinvestav.mx}, G.
Tavares--Velasco and J. J. Toscano \thanks{On leave from Facultad de 
Ciencias F\'{\i}sico Matem\'aticas, Universidad Aut\'onoma de Puebla.}
\thanks{E--mail: jtoscano@fcfm.buap.mx}}
\address{Departamento de F\'{\i}sica, CINVESTAV, Apartado Postal
14--740, 07000, M\'exico D. F.,M\'exico.}

\date{\today} 
\maketitle

\widetext

\begin{abstract}

A complete study of the one--loop induced decay $Z \to \bar{\nu}\nu
\gamma$ is presented within the framework of the Standard Model. The
advantages of using a nonlinear gauge are stressed. We have found that the
main contributions come from the electric dipole and the magnetic dipole
transitions of the $Z$ gauge boson and the neutrino, respectively.  We
obtain a branching ratio $B(Z \to \bar{\nu}\nu \gamma)=7.16\times
10^{-10}$, which is about four orders of magnitude smaller than the bound
recently obtained by the L3 Collaboration and thus it leaves open a window
to search for new physics effects in single--photon decays of the $Z$
boson.\protect\\

\end{abstract}


\section{Introduction} 
\label{int}

In the absence of any clear deviation from the Standard Model (SM) 
predictions, considerable work has been done recently to search for
processes that are forbidden or highly suppressed and may become a good
playground to test any new physics lying beyond the SM. The radiative
decay $Z \to \bar{\nu}\nu \gamma$ is a good example of this type of
processes because it arises at the one--loop level in the SM. The L3
Collaboration has searched for energetic photon events coming from
single--photon $Z$ decays. It was obtained a bound of one part in a
million for the respective branching ratio \cite{Acciarri}. Single photon
events in $Z$ decays have been associated to gauge boson and leptonic
substructure \cite{Barr,Boud}, supersymmetric particles \cite{Dicus}, a
possible neutrino magnetic moment \cite{Gould} and the production of a
Higgs boson with its associated CP--odd majoron \cite{Romao}. The decay $Z
\to \bar{\nu}\nu \gamma$ is expected also to be sensitive to the possible
anomalous electromagnetic properties of the $Z$ boson and the neutrinos. 
The latter point has become of increased interest after the recent
evidence of oscillations in atmospheric neutrinos \cite{Fukuda}. In a
recent study of the decay $Z \to \bar{\nu}\nu \gamma$ within the effective
Lagrangian approach \cite{Perez,Maltoni}, the L3 bound was used to get
direct constraints on dimension--six and dimension--eight operators
associated with the anomalous electromagnetic properties of the
neutrino.\protect\\

Though there is a great incentive in studying single--photon $Z$ decays,
to our knowledge its decay width has not been computed in the SM. We would
like to stress that in order to disentangle new contributions to this
process, it is important to count with a complete analysis of the SM
contributions. Our general aim is precisely to present explicit
expressions of a calculation of the SM decay $Z \to \bar{\nu}\nu \gamma$
at lowest order in perturbation theory. As we will see below, the
amplitude for this process can be written in terms of five independent
$\mathrm{U_e(1)}$ gauge structures.  Some of them can be identified in
terms of the so--called off--shell electromagnetic properties of the $Z$
gauge boson and the neutrino. It is well known that the $Z$ boson
possesses both an anapole moment (AM) and an electric dipole transition
(EDT) defined through the $ZZ\gamma^*$ and $ZZ^*\gamma$ couplings,
respectively (where $*$ stands for off--shell particles) \cite{Barr}. A
SM neutrino can have only off--shell electromagnetic properties
characterized by the $\nu\nu\gamma^*$ and $\nu\nu^*\gamma$ couplings. The
$\nu\nu\gamma^*$ coupling corresponds to an AM of the neutrino, which has
been extensively studied in the literature \cite{DSM,ZM}. It will be seen
below that the $\nu\nu^*\gamma$ coupling can be identified with a magnetic
dipole transition (MDT) of the neutrino. In this paper we have found that
the main contributions to the decay $Z \to \bar{\nu}\nu \gamma$ come
precisely from the MDT and EDT properties of the neutrino and the $Z$
boson, respectively. In particular, the MDT of the neutrino plays an
important role in this process.  Therefore, we will discuss with some
extent its $\mathrm{SU(2)}$ structure. For this purpose we will use a
nonlinear gauge which allows us to separate in a transparent way the
$\mathrm{SU(2)}$ and $\mathrm{U_e(1)}$ gauge structures and, as a
consequence, it will simplify considerably the calculation. In particular,
we have found that there is no $\mathrm{SU(2)}$ gauge link between the
$\nu\nu^*\gamma$ coupling and the box diagrams contributing to the decay.
This result should be contrasted with the situation found in the
calculation of the SM scattering amplitude of neutrinos with charged
leptons and nucleons. In this case, the $\mathrm{SU(2)}$ gauge structure
of the $\nu\nu\gamma^*$ coupling is directly linked to the box diagrams
contributing to the scattering process \cite{DSM}.\protect\\

In the nonlinear gauge the decay $Z \to \bar{\nu}\nu \gamma$ receives
contributions from reducible triangle diagrams characterized by the
$ZZ^*\gamma$ and $\nu\nu^*\gamma$ couplings, as well as from box diagrams
containing $W$ boson and lepton combined effects. It is well known that in
many four--body processes the box diagrams contributions can be neglected
compared with the triangle ones. However, this approximation fails when
the triangle and box diagrams possess common $\mathrm{U_e(1)}$ gauge
structures which induce form factors of the same order of magnitude. In
particular, the tensor structures arising from the box diagrams will have
a Lorentz covariant decomposition inducing the MDT of the neutrino. 
Therefore, it is not possible to neglect the box diagram contributions in
the decay $Z \to \bar{\nu}\nu \gamma$. Our numerical result confirms that
this decay is a SM highly suppressed process, about four orders of
magnitude below the L3 bound \cite{Acciarri}. Thus it leaves open a window
to search for new physics effects in single--photon events, mainly in the
neutrino sector.\protect\\

We have organized our presentation in the following way. In Sec. \ref{cal}
we present the analytic expressions for the $Z \to \bar{\nu}\nu \gamma$
decay width. In Sec.  \ref{num} we give details of the numerical analysis,
and the conclusions are presented in Sec. \ref{sum}.

\section{The nonlinear Gauge and the decay $Z \to \bar{\nu}\nu \gamma$}
\label{cal}

\begin{figure} 
\centerline{ \epsfysize=6cm \epsfbox{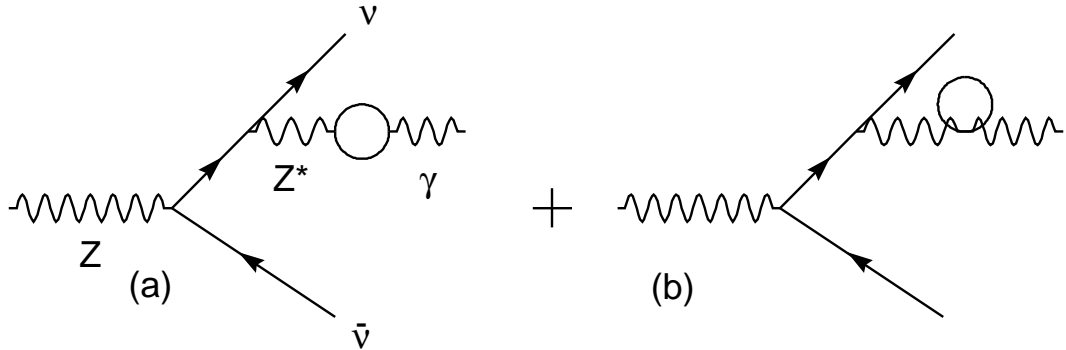} }
\caption{$Z^*-\gamma$ mixing term. Its contribution vanishes in the
nonlinear gauge.} 
\label{fig1} 
\end{figure}

\begin{figure}
\centerline{ \epsfysize=6cm \epsfbox{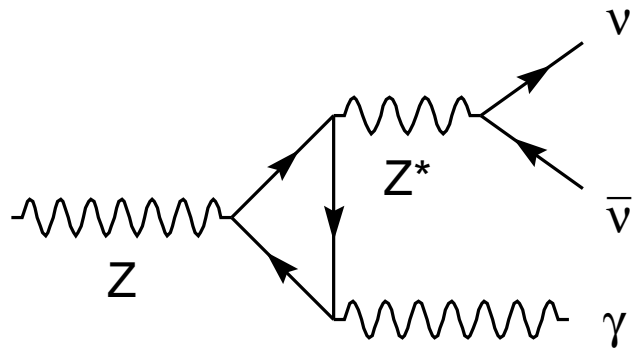} }
\caption{$Z Z^* \gamma$ coupling contribution. The crossed diagram must be
added.}
\label{fig2}
\end{figure}

The decay $Z \to \bar{\nu}\nu \gamma$ is a SM one--loop prediction which
receives contributions from all charged particles. Within the conventional
linear gauge, we can classify in five sets the Feynman diagrams
contributing to this decay at the one--loop level. The first three sets
are characterized by reducible diagrams including the $Z^*-\gamma$ mixing
term (Fig. \ref{fig1}) and the $ZZ^* \gamma$ and $\nu \nu^* \gamma$
couplings (Fig. \ref{fig2} and Fig. \ref{fig3}, respectively), while the
other two sets include box diagrams where the $Z$ boson couples to charged
leptons (Fig. \ref{fig4}) and to $W$ bosons (Fig.  \ref{fig5}).  For the
purpose of our calculation, it is suitable using a nonlinear gauge in
which there is no $\mathrm{SU(2)}$ gauge (nor $\mathrm{U_e(1)}$ gauge)
link between the $\nu\nu^*\gamma$ coupling and the box diagrams.  This
task is a difficult one to fulfill in a linear gauge, where
$\mathrm{SU(2)}$ gauge independence and $\mathrm{U_e(1)}$ gauge invariance
are only accomplished after adding up almost all the Feynman diagrams
involved in the calculation. Furthermore, in this type of gauges a
renormalization of the $Z^*-\gamma$ mixing term has to be done \cite{BR}.
In contrast, it can be shown that in the nonlinear gauge the amplitude for
this mixing term vanishes for any value of the $\mathrm{SU(2)}$ gauge
parameter $\xi$. We will show also that the $\nu\nu^*\gamma$ coupling is
by itself $\mathrm{SU(2)}$ gauge independent and $\mathrm{U_e(1)}$ gauge
invariant.\protect\\

A nonlinear $R_{\xi}$--gauge condition was first introduced by Fujikawa
\cite{fuj}. The ordinary derivative involved in the gauge--fixing term for
the $W$ boson is replaced by the respective $\mathrm{U_e(1)}$ covariant
derivative;  as a consequence, there are no $W^{\pm}G^{\mp}\gamma$
unphysical vertices, with $G^{\pm}$ the respective would-be Goldstone
boson. The $W$ boson and the charged ghosts sectors are separately
$\mathrm{U_e(1)}$ gauge invariant, leading to naive Ward identities. This
procedure was later extended to remove both $W^{\pm}G^{\mp}\gamma$ and
$W^{\pm}G^{\mp}Z$ vertices \cite{BD,M}. It is possible to proceed further
with this scheme and remove more unphysical vertices, such as
$H^0W^{\pm}G^{\mp}\gamma(Z)$ and $G^0W^{\pm}G^{\mp}\gamma(Z)$, with $G^0$
and $H^0$ the neutral would--be Goldstone and physical Higgs bosons,
respectively \cite{B,T}. For our present purposes, we will use a
gauge--fixing functional which is nonlinear in the vector sector but
linear in the scalar sector. \protect\\

The functionals which define the respective $W$, $Z$ and $A$ propagators
in the nonlinear gauge can be written as

\begin{eqnarray}
\label{eqgf}
f^+ &=& \bar{D}_{\mu}W^{+ \mu}-i {\xi}m_WG^+,\nonumber \protect\\
f^Z &=& \partial_{\mu}Z^{\mu}-{\xi}m_ZG^0,\nonumber \protect\\
f^A &=& \partial_{\mu}A^{\mu},
\end{eqnarray}

\noindent where $s_w=sin\theta_w$, $c_w=cos\theta_w$ and $\theta_w$ is the
weak mixing angle. We have also defined the following operator

\begin{equation}
\label{eqdb}
\bar{D}_{\mu} = \partial_{\mu}-ig^{\prime}B_{\mu},
\end{equation}

\noindent$Z_{\mu}=c_wW^3_{\mu}-s_wB_{\mu}$,
$A_{\mu}=s_wW^3_{\mu}+c_wB_{\mu}$, and $W^i_{\mu}$ and $B_{\mu}$ are the
gauge fields of the $\mathrm{SU(2)}$ and $\mathrm{U_Y(1)}$ groups,
respectively. We can see that the above operator contains the
electromagnetic covariant derivative, so that the $f^{\pm}$ functionals
transform covariantly under the $\mathrm{U_e(1)}$ group. The corresponding
gauge fixing Lagrangian is given by

\begin{equation}
\label{eqgfl}
{\cal L}_{GF} \ 
= \ -\frac{1}{\xi} \ f^+f^- \ - \ \frac{1}{2 \xi} \ (f^Z)^2 -
 \frac{1}{2 \xi} \ (f^A)^2,
\end{equation}

\noindent which removes the $W^{\pm}G^{\mp} \gamma$ and $W^{\pm}G^{\mp}Z$
vertices from the Higgs kinetic energy term
$(D_{\mu}\phi)^{\dag}(D^{\mu}\phi)$. The corresponding Faddeev--Popov
Lagrangian is given by

\begin{eqnarray}
\label{eqfpl}
{\cal L}_{FP} &=& -\bar{c}^-[\bar{D}_{\mu}\hat{D}^{\mu}+
{\xi}m_W(m_W+H^0+iG^0)]c^+ -igc_w\bar{c}^-
\bar{D}_{\mu}(W^{+\mu}c_Z)-ie\bar{c}^-
\bar{D}_{\mu}(W^{+\mu}c_{\gamma}) \nonumber \protect\\
&-& i\frac{gs^{2}_w}{c_w}W^{+\mu}\bar{c}^-
(\partial^{\mu}c_Z)+ieW^+_{\mu}\bar{c}^-
(\partial^{\mu}c_{\gamma}) -igc_wW^{+\mu}(\partial_{\mu}
\bar{c}_Z)c^--ieW^{+\mu}(\partial_{\mu}\bar{c}_{\gamma})c^- \nonumber 
\protect\\ &-& gc_{2w}m_Z\xi(G^+\bar{c}^-c_Z+G^+\bar{c}_Zc^-)-em_W\xi(G^+
\bar{c}^-c_{\gamma}+G^+\bar{c}_{\gamma}c^-)+H.C. \nonumber \protect\\
&-& \bar{c}_Z[\Box+{\xi}m_Z(m_Z+H^0)]c_Z-\bar{c}_{\gamma}{\Box}c_{\gamma},
\end{eqnarray}

\noindent where $\bar{c}^{\pm}(c^{\pm})$, $\bar{c}_Z(c_Z)$, and
$\bar{c}_{\gamma}(c_{\gamma})$ are the pairs of ghosts associated with the
$W^{\pm}$, $Z$, and $A$ gauge bosons, respectively, and
$c_{2w}=c^2_w-s^2_w$. The charged ghosts satisfy
$(\bar{c}^+)^{\dag}=\bar{c}^-$ and $(c^+)^{\dag}=c^-$. We have introduced
the following operator

\begin{equation}
\label{eqdw}
\hat{D}_{\mu} = \partial_{\mu}-igW^3_{\mu},
\end{equation}

\noindent which also contains the electromagnetic covariant derivative
\footnote{Notice that if we use in (\ref{eqgf}) $\hat{D}_{\mu}$ instead of
$\bar{D}_{\mu}$, as it was pointed out in \cite{D}, the
$W^{\pm}G^{\mp}\gamma$ vertex is removed but the $W^{\pm}G^{\mp}Z$ vertex
only becomes modified.}. Therefore, just as it was the case with the
$f^-f^+$ term in ${\cal L}_{GF}$, the ${\cal L}_{FP}$ Lagrangian is
invariant under the $\mathrm{U_e(1)}$ group. Since both the Yang--Mills
and the kinetic energy Higgs sectors can be expressed in terms of the
$\bar{D}_{\mu}$ and $\hat{D}_{\mu}$ operators, all the charged sector of
the SM satisfies naive Ward identities.\protect\\
 
To calculate the amplitude for the decay $Z \to \bar{\nu}\nu \gamma$ in
the nonlinear gauge, the Feynman rules for the
$A_{\lambda}(k_1)W^+_{\rho}(k_2)W^-_{\eta}(k_3)$,
$Z_{\lambda}(k_1)W^+_{\rho}(k_2)W^-_{\eta}(k_3)$, and
$A_{\rho}Z_{\lambda}W^+_{\alpha}W^-_{\beta}$ vertices are required

\begin{eqnarray}
\label{eqver}
{\Gamma}^{AWW}_{\lambda\rho\eta} 
&=& -ie[(k_3-k_1+\frac{k_2}{\xi})_{\rho}g_{\lambda\eta}
+(k_1-k_2-\frac{k_3}{\xi})_{\eta}g_{\lambda\rho}+
(k_2-k_3)_{\lambda}g_{\rho\eta}], \nonumber \protect\\
{\Gamma}^{ZWW}_{\lambda\rho\eta} &=& 
-igc_w[(k_3-k_1-\frac{s^2_w}{c^2_w}\frac{k_2}{\xi})_{\rho}g_{\lambda\eta}
+(k_1-k_2+\frac{s^2_w}{c^2_w}\frac{k_3}{\xi})_{\eta}g_{\lambda\rho}
+(k_2-k_3)_{\lambda}g_{\rho\eta}], \nonumber \protect\\
{\Gamma}^{AZWW}_{\rho\lambda\alpha\beta} 
&=& -iegc_w[2g_{\alpha\beta}g_{\rho\lambda}-(1
+\frac{1}{\xi}\frac{s^2_w}{c^2_w})(g_{\alpha\lambda}g_{\beta\rho} +
g_{\beta\lambda}g_{\alpha\rho})],
\end{eqnarray}

\noindent where all the momenta are incoming. We have taken the lepton
masses to be zero except where large logs (collinear singularities) arise. 
This approximation and the use of the nonlinear gauge reduce considerably
the number of diagrams involved in the calculation. We have calculated the
Feynman amplitudes using the computer program FeynCalc
\cite{Mer91}.\protect\\

In the linear gauge the $Z^*-\gamma$ mixing term receives contributions
from $W^{\pm}$ gauge bosons, charged ghosts and $W^{\pm}G^{\mp}$ combined
effects \footnote{Since the Lagrangian which defines the $G^{\pm}$ boson
is not modified in its interactions with the photon by the gauge fixing
procedure used for the $\mathrm{SU(2)}$ group, and since it resembles
scalar electrodynamics, the contribution of this field to the $Z^*-\gamma$
mixing term vanishes in any gauge.}. In contrast, in the nonlinear gauge
the $W^{\pm}G^{\mp}$ combined effects are absent.  From (\ref{eqgf}),
(\ref{eqgfl}) and (\ref{eqfpl}) we can see that the $f^+f^-$ term in the
${\cal L}_{GF}$ and the ${\cal L}_{FP}$ Lagrangians also presents explicit
$\mathrm{U_e(1)}$ gauge invariance.  Consequently, in this gauge we expect
a zero contribution arising from all charged particles. This fact was
shown before in the Feynman--t'Hooft version of the nonlinear gauge
\cite{T,D}. We will show that this result is valid for any value of the
gauge parameter $\xi$. Notice that the charged part of the ${\cal L}_{FP}$
Lagrangian resembles scalar electrodynamics (see (\ref{eqfpl})), so that
we do not expect anything new coming from these particles in the general
nonlinear $R_{\xi}$--gauge.  In contrast, both the $W$--propagator and the
$W^{\pm}W^{\mp}\gamma(Z)$ vertices have a nontrivial dependence on the
$\xi$ parameter. The corresponding amplitude which arises from the
diagrams shown in Figs. \ref{fig1}(a) and \ref{fig1}(b) is given by

\begin{equation}
\label{eqmt1}
\Pi^{Z^*\gamma}_{\alpha\beta}=
\frac{iegc_wm^2_W}{64\pi^2}[U_a+U_b+
\frac{s^2_w}{c^2_w}(V_a+V_b)]g_{\alpha\beta},
\end{equation}

\noindent with $U_a(U_b)$ and $V_a(V_b)$ characterizing the loops in Figs. 
\ref{fig1}(a) (\ref{fig1}(b)).  After a straightforward calculation, we
have found that there is not $Z^*-\gamma$ mixing term in the general
nonlinear $R_\xi$--gauge

\begin{eqnarray}
\label{eqmt2}
U_a &=& -U_b=
\xi^2[5+6B_0(0,\xi m^2_W,\xi m^2_W)]-3[1+6B_0(0,m^2_W,m^2_W)], 
\nonumber \protect\\ V_a &=& -V_b=
-\xi[3+2B_0(0,\xi m^2_W,\xi m^2_W)]-\frac{1}{\xi}[5+6B_0(0,m^2_W,m^2_W)],
\end{eqnarray}

\noindent where $B_0(0,\xi m^2_W,\xi m^2_W)$ and $B_0(0,m^2_W,m^2_W)$ are
Passarino--Veltman two--point scalar functions \cite{Pas79}.\protect\\

In this manner, the width for the decay $Z \to \bar{\nu}\nu \gamma$ can be
expressed as

\begin{equation}
\label{eqwidth}
\Gamma(Z \to \bar{\nu}\nu\gamma) =\frac{m_Z}{256 \pi^3}\int_{0}^{1}dx
\int_{0}^{1-x}dy\arrowvert \bar{\cal M}\arrowvert^2,
\end{equation}

\noindent where $\cal M$ is the invariant amplitude

\begin{equation}
\label{eqamp}
{\cal M}=
\bar{u}_L(p_1)[{\cal M}^{\alpha\beta}_{Z^*}+
{\cal M}^{\alpha\beta}_{\nu^*}+{\cal M}^{\alpha\beta}_{Box4}+
{\cal M}^{\alpha\beta}_{Box5}]v_R(p_2){\epsilon}_{\alpha}
(k_2,{\lambda}_2){\epsilon}^*_{\beta}(k_1,{\lambda}_1), \protect\\
\end{equation}

\noindent and the ${\cal M}^{\alpha\beta}_{Z^*}$, ${\cal
M}^{\alpha\beta}_{\nu^*}$, ${\cal M}^{\alpha\beta}_{Box4}$, ${\cal
M}^{\alpha\beta}_{Box5}$ amplitudes correspond to the respective diagrams
shown in Figs. \ref{fig2}--\ref{fig5}. The explicit form of these
amplitudes will be given below. Moreover, $p_1$($p_2$) are the
neutrino(antineutrino)  momenta, $k_1$($k_2$) are the $\gamma$($Z$) 
momenta, ${\epsilon}_{\beta}$(${\epsilon}_{\alpha}$) are the $\gamma$($Z$) 
polarization vectors, and $u_L$=$\frac{1}{2}(1-{\gamma}_5)u$. It is useful
to define the following dimensionless quantities

\begin{equation}
\label{eqdefin}
x=\frac{2k_1.p_1}{m^2_Z},\hspace{5mm} y=\frac{2k_1.p_2}{m^2_Z}.
\end{equation}

\noindent and the combinations $\delta=1-x-y$, $\delta_+=x+y$,
$\delta_-=x-y$, $\delta_x=1-x$, $\delta_y=1-y$.

\subsection{$ZZ^* \gamma$ coupling contribution}
\label{zzp}

The $ZZ^*\gamma$ coupling contributes to the decay $Z \to \bar{\nu}\nu
\gamma$ through the reducible diagram shown in Fig.  \ref{fig2}. The
contributions to this coupling are due to the fermion triangle diagram
responsible for the ABJ anomaly, which of course vanishes with the SM
fermion multiplet assignment. The corresponding amplitude must be
$\mathrm{U_e(1)}$ gauge invariant by itself. This amplitude was obtained
by Barroso et al. some time ago\cite{Barr}. We have reproduced this result
to compute the corresponding contribution to the decay $Z \to \bar{\nu}\nu
\gamma$.  Since one $Z$ boson and the photon are on--shell, we take
$k^{\beta}_1{\epsilon}_{\beta}=k^{\alpha}_2{\epsilon}_{\alpha}=0$,
$k^2_1=0$, $k^2_2=m^2_Z$. In addition, using Bose statistics and the Ward
identity satisfied by the electromagnetic current, we obtain for this
amplitude

\begin{eqnarray}
\label{eqzzp1}
G^{\alpha\beta\gamma} 
&=& \frac{ig^2e}{2c^2_w{\pi}^2} \ \sum_f g_V^f g_A^f Q_f \ 
\lbrace \ \frac{1}{q^2-m^2_Z}[ m^2_fC_0(m^2_Z,q^2) \nonumber \protect\\
&-& \frac{m^2_Z}{2(q^2-m^2_Z)}(B_0(m^2_Z)-B_0(q^2))+ \frac{1}{2}]\epsilon^
{\alpha\beta\lambda\rho}k_{2\lambda}q_{\rho}q^{\gamma}  \nonumber \protect\\
& +& \frac{1}{2}\frac{q^2+m^2_Z}{q^2-m^2_Z} [ m^2_fC_0(m^2_Z,q^2)-
\frac{q^2m^2_Z}{q^4-m^4_Z}(B_0(m^2_Z)-B_0(q^2))+\frac{1}{2} ] 
\epsilon^{\alpha\beta\gamma\lambda}k_{1\lambda}  \rbrace, 
\end{eqnarray}

\noindent where $C_0(m^2_Z,q^2)=C_0(0,m^2_Z,q^2,m^2_f,m^2_f,m^2_f)$,
$B_0(m^2_Z)=B_0(m^2_Z,m^2_f,m^2_f)$, and $B_0(q^2)=B_0(q^2,m^2_f,m^2_f)$
are Passarino--Veltman three-- and two--point scalar functions written in
the notation of \cite{Mer91}. $Q_f$ is the electric charge in units of
$e$, $q^2=(p_1+p_2)^2=m^2_Z{\delta}$, $g^f_V=t^f_3-2Q^fs^2_w$, and
$g^f_A=t^f_3$, with $t^f_3$ the third component of weak isospin.\protect\\

Now, it is straightforward to obtain the respective contribution to the
decay $Z \to \bar{\nu}\nu \gamma$

\begin{equation}
\label{eqzzp3}
{\cal M}^{\alpha\beta}_{Z^*}=
\frac{2i{\alpha}^2}{(s_wc_w)^3}{\gamma}_{\lambda}\sum_{f}g^f_V
g^f_AQ^fF(m^2_Z,q^2,m^2_f){\epsilon}^{\alpha\beta\lambda\rho}k_{1\rho},
\end{equation}

\noindent where

\begin{equation}
\label{eqzzp4}
F(m^2_Z,q^2,m^2_f)=
\frac{1}{(q^2-m^2_Z)^2}[(q^2+m^2_Z)m^2_f
C_0(m^2_Z,q^2)-\frac{q^2m^2_Z}{q^2-m^2_Z}(B_0(m^2_Z)-B_0(q^2))]. 
\end{equation}

In addition, we have removed from (\ref{eqzzp4}) the mass independent term
which vanish when we perform the summation over fermion
families.\protect\\

As it was shown in \cite{Barr}, and more recently in \cite{BH}, the $Z$
boson possesses only off--shell electromagnetic properties characterized
by an electric dipole transition (EDT) and an anapole moment (AM), which
correspond to the $ZZ^*\gamma$ and $ZZ{\gamma}^*$ couplings, respectively. 
Since the on--shell $ZZ\gamma$ vertex is zero, the $Z$ boson does not
possess on--shell (physical) electromagnetic properties. So the
contribution of the $ZZ^*\gamma$ coupling to the decay $Z \to \bar{\nu}\nu
\gamma$ corresponds to an EDT, which is a transfer--momentum dependent
quantity. This EDT electromagnetic property of the $Z$ boson depends only
on the number of families. Indeed, as we will discuss in the following
section, the contributions arising from the first two families, which are
nearly mass degenerated, are very small compared with the corresponding
contribution of the third family, where the masses involved have an
important gap.\protect\\
 
\subsection{$ \nu \nu^* \gamma$ coupling contribution} 
\label{nnp}

\begin{figure} 
\centerline{ \epsfysize=6cm \epsfbox{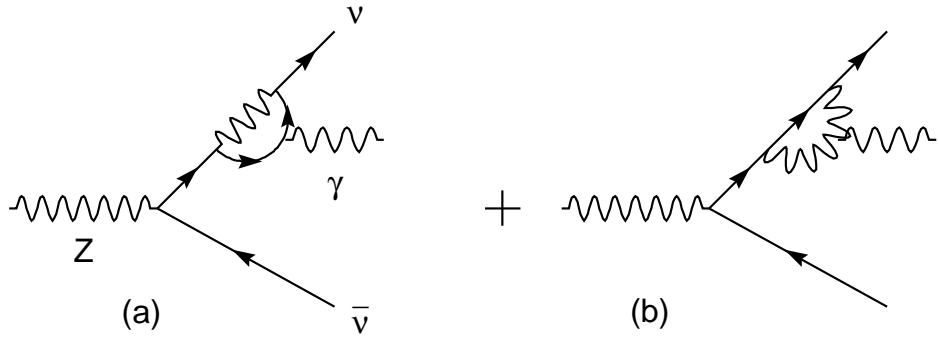} }
\caption{$\nu\nu^* \gamma$ coupling contribution. The corresponding
diagram with the loop in $\bar\nu$ must be added.} 
\label{fig3}
\end{figure}

The decay $Z \to \bar{\nu}\nu \gamma$ receives contributions coming from
the $ \nu \nu^* \gamma$ coupling through the diagrams shown in Figs.
\ref{fig3}(a) and \ref{fig3}(b), where the photon couples to leptons and
to $W$ bosons, respectively. We will present the calculation for the $ \nu
\nu^* \gamma$ coupling in the general nonlinear $R_\xi$--gauge and then we
will use it to give the corresponding amplitude for the decay $Z \to
\bar{\nu}\nu \gamma$. The triangle \ref{fig3}(a)  depends on the gauge
parameter $\xi$ only through the $W$--propagator. On the other hand, the
triangle \ref{fig3}(b) has a more complicated $\xi$ dependence through
both the $W$--propagator and the $W^{\pm}W^{\mp}\gamma$ vertex.  Taking
into account that a neutrino and the photon are on--shell, the amplitude
for the $ \nu \nu^* \gamma$ coupling can be written as

\begin{equation}
\label{eqnnp1}
\Gamma^{\nu\nu^*\gamma}_{\beta}
=\frac{ig^2e}{16\pi^2}\frac{1}{p^2}[\frac{p^2}{2}\gamma_{\beta}(F_a+F_b)
+p_{1\beta}\FMSlash{k}_1 (G_a+G_b) ],
\end{equation}

\noindent where $p^2=(k_1+p_1)^2=xm^2_Z$ is the momentum transferred by
the $Z$ boson. $F_a(F_b)$ and $G_a(G_b)$ are the amplitudes corresponding
to diagrams \ref{fig3}(a) (\ref{fig3}(b))

\begin{eqnarray}
\label{eqnnp2}
F_a &=&\frac{m^2_W}{p^2}+\frac{1}{2}(1-\xi)+\frac{1}{2}(1-\frac{2m^2_W}{p^2})
(1-\frac{p^2}{m^2_W})B_0(p^2,0,m^2_W) \nonumber \protect\\
&-&\frac{\xi}{2}B_0(0,\xi m^2_W,\xi m^2_W) +\frac{1}{2}(\frac{p^2}{m^2_W}-\xi)
B_0(p^2,0,\xi m^2_W) \nonumber \protect\\
&-&(\frac{3}{2}-\frac{m^2_W}{p^2})B_0(0,m^2_W,m^2_W),
\end{eqnarray}

\begin{equation}
\label{eqnnp3}
G_a=1-\frac{2m^2_W}{p^2}-2(1-\frac{m^2_W}{p^2})[B_0(p^2,0,m^2_W)-
B_0(0,m^2_W,m^2_W)],
\end{equation}

\begin{eqnarray}
\label{eqnnp4} 
F_b &=& 2-\frac{m^2_W}{p^2}-\frac{1}{2}(1-\xi)-\frac{1}{2}(1-\frac{2m^2_W}
{p^2})(1-\frac{p^2}{m^2_W})B_0(p^2,0,m^2_W)+\frac{1}{2}\xi 
B_0(0,\xi m^2_W,\xi m^2_W) \nonumber \protect\\
&-& \frac{1}{2}(\frac{p^2}{m^2_W}-\xi)B_0(p^2,0,\xi m^2_W)+(\frac{3}{2}-
\frac{m^2_W}{p^2})B_0(0,m^2_W,m^2_W)+2(m^2_W-p^2)C_0(3), \nonumber \protect\\
\end{eqnarray}

\begin{equation}
\label{eqnnp5}
G_b=-3+\frac{2m^2_W}{p^2}+2(1-\frac{m^2_W}{p^2})[B_0(p^2,0,m^2_W)-
B_0(0,m^2_W,m^2_W)]-2(m^2_W-p^2)C_0(3), \nonumber \protect\\
\end{equation}

\noindent where the three--point scalar function is defined in the
Appendix. From (\ref{eqnnp1})--(\ref{eqnnp5}) it is evident that the
amplitude for the $\nu\nu^*\gamma$ coupling is a $\mathrm{SU(2)}$ gauge
independent and $\mathrm{U_e(1)}$ gauge invariant quantity that can be
written as

\begin{equation}
\label{eqnnp6}
\Gamma^{\nu\nu^*\gamma}_{\beta}=i\frac{A(p^2)}{p^2}k^{\mu}_1\sigma_{\mu\beta}
\FMSlash{p},
\end{equation}

\noindent where
$\sigma_{\mu\beta}=\frac{i}{2}(\gamma_{\mu}\gamma_{\beta}-\gamma_{\beta}
\gamma_{\mu})$, and

\begin{equation}
\label{eqnnp7}
A(p^2)=\frac{g^2e}{16\pi^2}[1+(m^2_W-p^2)C_0(3)].
\end{equation}

It is well known that in the SM the neutrino have no static
electromagnetic properties. It is easy to see that our result
(\ref{eqnnp7}) is consistent with this fact by noting that the $A(p^2)$
form factor vanishes when the neutrino is on--shell. Using the above
results, it is straightforward to calculate the corresponding amplitude
for the decay $Z \to \bar{\nu}\nu \gamma$. Taking into account all
diagrams shown in Fig.  \ref{fig3}, we have

\begin{equation}
\label{eqnnp8}
{\cal M}^{\alpha\beta}_{\nu^*}=\frac{\alpha^2}{2c_ws^3_wm^2_Z}[A_1(x)
\Gamma^{\alpha\beta}_1+A_2(y)\Gamma^{\alpha\beta}_2],
\end{equation}

\noindent where the form factors $A_1(x)$ and $A_2(y)$ are
transfer--momentum dependent quantities given by

\begin{eqnarray} \label{eqnnp9}
A_1(x)=\frac{1}{x}[1+m^2_Z(c^2_w-x)C_0(3)], \ \
A_2(y)=A_1(x\longrightarrow y).  \end{eqnarray}

\noindent and the $\mathrm{U_e(1)}$ gauge structures are given by

\begin{eqnarray}
\label{eqnnp10}
\Gamma^{\alpha\beta}_1=ik_{1\mu}\sigma^{\beta\mu}\gamma^{\alpha}, \ \ 
\Gamma^{\alpha\beta}_2=ik_{1\mu}\gamma^{\alpha}\sigma^{\mu\beta}.
\end{eqnarray}

The ${\cal M}^{\alpha\beta}_{\nu^*}$ amplitude can be interpreted in terms
of the off--shell electromagnetic properties of the neutrino. In fact, we
can identify in the gauge structure of this amplitude the familiar form of
a fermionic magnetic dipole transition. Since the corresponding form
factor is an on--shell vanishing quantity, we can relate it with a MDT of
the neutrino. We would like also to stress an aspect related with the
off--shell electromagnetic properties of the neutrino, namely, the MDT an
AM. We have found that the MDT is an $\mathrm{SU(2)}$ gauge independent
and $\mathrm{U_e(1)}$ gauge invariant quantity. This result must be
contrasted with the one obtained for the AM arising from the
$\bar{\nu}\nu\gamma^*$ coupling which depends on the gauge parameter $\xi$
\cite{DSM,ZM}.\protect\\
 
\subsection{Box diagrams contribution}
\label{znnp}

In this section we present the amplitudes for the box diagrams
contributing to the decay $Z \to \bar{\nu}\nu \gamma$. These diagrams,
shown in Figs. \ref{fig4} and \ref{fig5}, induce terms with the same
Lorentz structure of the neutrino MDT. This implies that we can not
neglect these contributions because they are of the same order of
magnitude than the one induced by the $\nu\nu^*\gamma$ coupling. We have
found that there is not $Z^*-\gamma$ mixing term in the general nonlinear
$R_\xi$--gauge. Also, we have found that the $\nu\nu^*\gamma$ coupling is
$\mathrm{SU(2)}$ gauge independent and $\mathrm{U_e(1)}$ gauge invariant.
This implies that the contributions arising from the box diagrams must
respect these properties by themselves.  Even more, since the two sets of
box diagrams arise from different $Z$ boson couplings, we expect that they
are separately $\mathrm{SU(2)}$ gauge independent and $\mathrm{U_e(1)}$
gauge invariant. We have found that in the Feynman--t'Hooft version of the
nonlinear $R_{\xi}$--gauge the respective amplitude is $\mathrm{U_e(1)}$
invariant and finite for each set of diagrams

\begin{equation}
\label{eqab4}
{\cal M}^{\alpha\beta}_{Box4}=
\frac{\alpha^2 c_{2w}}{2c_ws^3_wm^2_Z}
\sum^7_{i=1}A^l_i(x,y)\Gamma^{\alpha\beta}_i
\end{equation}

\noindent and

\begin{equation}
\label{eqab5}
{\cal M}^{\alpha\beta}_{Box5}=\frac{\alpha^2c_w}{s^3_wm^2_Z}\sum^7_{i=1}
A^W_i(x,y)\Gamma^{\alpha\beta}_i,
\end{equation}

\noindent where the $\Gamma^{\alpha\beta}_1$ and $\Gamma^{\alpha\beta}_2$
$\mathrm{U_e(1)}$ gauge structures are given in (\ref{eqnnp10}) and

\begin{eqnarray}
\label{eqgs}
\Gamma^{\alpha\beta}_3 &=& 
\frac{1}{m^2_Z}\gamma^{\alpha}
(k_1\cdot p_1p^{\beta}_2-k_1\cdot p_2p^{\beta}_1), \nonumber \protect\\
\Gamma^{\alpha\beta}_4 &=& 
\frac{1}{m^2_Z}\gamma_{\mu}p^{\alpha}
_1(k_1\cdot p_1g^{\mu\beta}-k^{\mu}_1p^{\beta}_1), 
\ \ \Gamma^{\alpha\beta}_6=\Gamma^{\alpha\beta}_4(p_1\leftrightarrow p_2), 
\nonumber \protect\\
\Gamma^{\alpha\beta}_5 &=& 
\frac{1}{m^2_Z}\gamma_{\mu}p^{\alpha}_2(k_1\cdot p_1g^{\mu\beta}-
k^{\mu}_1p^{\beta}_1), \ \ \Gamma^{\alpha\beta}_7=
\Gamma^{\alpha\beta}_5(p_1\leftrightarrow p_2),
\end{eqnarray}

\noindent while the $A^l_i$ and $A^W_i$ functions are given by 

\begin{eqnarray}
\label{eqal1}
A^l_1(x,y) &=& \frac{1}{2xy\delta_+}\lbrace\delta_-[B_0(5)-B_0(6)]-x\delta_+
[B_0(6)-B_0(1)]-y[B_0(6)-B_0(2)]\rbrace \nonumber \protect\\ 
&+& \frac{m^2_Z}{4\delta x^2y^2}\lbrace\delta_-\delta_+[x(x+2\delta_y)y^2+
2xy(1+x)c^2_w+\delta_-\delta_+c^4_w]C_0(3) \nonumber \protect\\
&+& y\delta[x(5x+2\delta_y)y^2+2xy(2\delta_+-\delta_x)c^2_w+\delta_-\delta_+
c^4_w]C_0(4) \nonumber \protect\\
&+& x[x\delta(x\delta-2y^2)-2x(x-\delta_-\delta_+)c^2_w-\delta_-\delta_+c^4_w]
C_0(7) \nonumber \protect\\
&+&y[\delta^2y^2-2y(x\delta_x+y\delta_y)c^2_w-\delta_-\delta_+c^4_w]C_0(8) 
\nonumber \protect\\
&+&x[y\delta_x(\delta^2_x+2(\delta_x+2y))+2(\delta^2_x+xy\delta-y\delta_y)
c^2_w]C_0(9) \nonumber \protect\\
&+& y\delta_y[y\delta_y(4x-\delta_y)+2(x-y\delta_y)c^2_w]C_0(10)+\delta^2
[\delta\delta_-\delta_++2xy^2+\delta_-\delta_+c^2_w]C_0(11)\rbrace 
\nonumber \protect\\
&+& \frac{m^4_Z}{4\delta x^2y^2}\lbrace[(x+2\delta_y)x^2y^3-xy^2(x(3\delta+10)
-2y\delta_y)c^2_w+xy(\delta^2+2x(1+3y)+4\delta_y)c^4_w \nonumber \protect\\
&-&\delta_-\delta_+c^6_w]D_0(4) +(x\delta+\delta_+c^2_w)[x\delta(x\delta-2y^2)
+2x(x-\delta_-\delta_+)c^2_w+\delta_-\delta_+c^4_w]D_0(5) \nonumber \protect\\
&-& (y\delta+\delta_+c^2_w)[y^2\delta^2-2y(\delta^2_+-y)c^2_w+\delta_-\delta_
+c^4_w]D_0(6)\rbrace,
\end{eqnarray} 

\begin{equation}
\label{eqal2}
A^l_2(x,y)=-A^l_1(x\leftrightarrow y),
\end{equation}

\begin{eqnarray}
\label{eqal3}
A^l_3(x,y) &=& \frac{1}{2xy\delta}\lbrace[B_0(6)-B_0(5)]+2xy[B_0(1)-B_0(6)]
\rbrace \nonumber \protect\\
&+& \frac{m^2_Z}{2x^2y^2\delta^2}\lbrace x[xy(3xy+2\delta^2)+4xyc^2_w+
\delta^2_+c^2_w]C_0(3) \nonumber \protect\\
&+& x[x(x+2y)\delta^2-2x(x^2+y(y-2\delta_+))c^2_w+\delta^2_+c^4_w]C_0(7) 
\nonumber \protect\\
&-& x[x^2(3y^2\delta_x+y^3)+2x(y^2+x\delta^2)]C_0(9)+x^2\delta^2(\delta+
2c^2_w)C_0(11)\rbrace \nonumber \protect\\
&+& \frac{m^4_Z}{2x^2y^2\delta^2}\lbrace
[\frac{1}{2}xy(\delta^2-3xy)+(2x^2(x-2)+xy+
\frac{1}{2}\delta(2+3xy))c^2_w+\frac{1}{2}\delta(\delta-6)c^4_w]
D_0(4) \nonumber \protect\\
&-& (x\delta+\delta_+c^2_w)[x\delta^2(x+2y)+2x(\delta_x(x+2y)-y^2)c^2_w+
\delta^2_+c^4_w]D_0(5)\rbrace +(x\leftrightarrow y),
\end{eqnarray}

\begin{eqnarray}
\label{eqal4}
A^l_4(x,y) &=& \frac{2}{x\delta_+}+\frac{2}{x^2\delta_y\delta^2_+}
\lbrace(2x+y)[B_0(5)-B_0(6)+y(B_0(6)-B_0(2))]+x^2[B_0(5)-B_0(2)]\rbrace 
\nonumber \protect\\
&+& \frac{m^2_Z}{x^3y\delta^2}\lbrace x[x(x^2\delta-2\delta^2)+yc^2_w]
C_0(3)+y(x\delta^2_x+yc^2_w)C_0(4) \nonumber \protect\\
&+& xyc^4_wC_0(7)+y^2(\delta+c_w)^2C_0(8)-x\delta^3_xc^2_wC_0(9) 
\nonumber \protect\\
&+& [-y\delta_y\delta^2+\delta_y(x-2\delta_-\delta_++2y\delta_x)c^2_w+
\frac{x^3}{\delta_y}(y(2+y)-1)c^2_w]C_0(10) \nonumber \protect\\
&+& y\delta^2(\delta+2c^2_w)C_0(11)\rbrace \nonumber \protect\\
&+& \frac{m^4_Z}{x^3y\delta^2}\lbrace [x^2y\delta^2_x-x(x\delta^2_x+
xy(2+x)-y(3-y))c^2_w+y\delta_+c^4_w]D_0(4) \nonumber \protect\\
&+& y(x\delta+\delta_+c^2_w)[x(x+2y)\delta^2+2x(\delta_x(x+2y)-y^2)c^2_w+
\delta^2_+c^4_w]D_0(5)+y(y\delta+\delta_+c^2_w)(\delta+c^2_w)^2D_0(6)\rbrace,
\end{eqnarray}

\begin{eqnarray}
\label{eqal5}
A^l_5(x,y) &=& \frac{2}{x\delta_+}+\frac{1}{x^2y\delta^2\delta^2_+}
\lbrace [y\delta_+(x-2\delta_y)-xy\delta+x^2\delta_+][B_0(6)-B_0(5)]+x^2
\delta^2_+[B_0(6)-B_0(1)] \nonumber \protect\\
&+& y\delta^2_+(x-2\delta_y)[B_0(2)-B_0(6)]\rbrace \nonumber \protect\\
&+& \frac{m^2_Z}{2x^3y^2\delta^2}\lbrace [x(-3x^2y^2-2xy(x(2x+\delta_x))-
\delta)c^2_w-(x\delta_-\delta_++2y^2\delta)c^4_w]C_0(3) \nonumber \protect\\
&-& [3x^3y^2+2xy(\delta\delta_x+2x^2)c^2_w+(x\delta_-\delta_++2y^2
\delta)c^4_w]C_0(4) \nonumber \protect\\
&+& x[-x^3\delta^2+2x(x(\delta_-\delta_+-x)+2y^2\delta)c^2_w-(x\delta_-
\delta_++2y^2\delta)c^4_w]C_0(7) \nonumber \protect\\
&+& y[y^2\delta^2(x-2\delta)+2y(x^2\delta_+-y(x\delta-2\delta^2))c^2_w-
(x\delta_-\delta_++2y^3\delta)c^4_w]C_0(8) \nonumber \protect\\
&+& x[x^2\delta_x(3y^3+\delta^2)+2(x^2(y\delta_x+\delta^2)+y
\delta(1-2x))c^2_w]C_0(9) \nonumber \protect\\
&+& y\delta_y[y(x(3x^2-\delta^2)+2\delta^3)+2(x(x^2-\delta^2)+2y
\delta^2)c^2_w]C_0(10) \nonumber \protect\\
&-& \delta^2[\delta(x\delta_-\delta_++2y^2\delta)+2(xy\delta_++2y^2
\delta)c^2_w]C_0(11)\rbrace \nonumber \protect\\
&+& \frac{m^4_Z}{2x^3y^2\delta^2}\lbrace [3x^4y^3+x^2y^2(-2\delta(y+3
\delta)+x(5y+66)+7x^2)c^2_w \nonumber \protect\\
&+& xy(6y\delta^2+x(y(y-4\delta)-2\delta^2)+2x^2(3y+\delta)+5x^3)c^4_w+
(x\delta_-\delta_++2y^2\delta)c^6_w]D_0(4) \nonumber \protect\\
&+& [(x\delta+\delta_+c^2_w)(x^3\delta^2+2x(x(x\delta_x+3y^2)-2y^2
\delta_y))c^2_w+(2y\delta_y+x(x^2-3y^2))c^4_w]D_0(5) \nonumber \protect\\
&+& [y^2\delta^2(2\delta_y-3x)+2y(y\delta_y(2\delta_y-5x)+x^3)c^2_w+((2
\delta_y-y)y^2-3xy^2)c^4_w]D_0(6)\rbrace,
\end{eqnarray}

\begin{equation}
\label{eqal6}
A^l_6(x,y)=-A^l_4(x\leftrightarrow y),
\end{equation}

\begin{equation}
\label{eqal7}
A^l_7(x,y)=-A^l_5(x\leftrightarrow y),
\end{equation}

\begin{eqnarray}
\label{eqaw1}
A^W_1(x,y) &=& \frac{1}{2xy\delta_+}\lbrace \delta_-[B_0(4)-B_0(3)]+\delta_+
[x(B_0(1)-B_0(4))-y(B_0(4)-B_0(2))]\rbrace \nonumber \protect\\
&+& \frac{m^2_Z}{4x^2y^2\delta}\lbrace [\delta\delta_+(-x^2\delta+y^2(7x+y
\delta))-2(\delta\delta_-(x^2+y^2)+4xy\delta^2_+)c^2_w-2\delta_-\delta^2_+
c^4_w]C_0(1) \nonumber \protect\\
&-& \delta^2[x^2(x-\delta_y)+y^2(3x+\delta_y)+2\delta_-\delta_+c^2_w]
C_0(2) \nonumber \protect\\
&+& x[x\delta(x\delta+2y^2)-2x(\delta_-+\delta^2_+)c^2_w+\delta_-\delta_+
c^4_w]C_0(3) \nonumber \protect\\
&-& y[y^2\delta(5x+\delta_y)+2y(\delta_-(1+\delta_-)-4x^2)c^2_w+
\delta_-\delta_+c^4_w]C_0(4) \nonumber \protect\\
&-& x[1-3y^2-\delta_+(2-\delta_+)-2(x+y(xy-\delta_y)-x^2(2-
\delta_+))c^2_w]C_0(5) \nonumber \protect\\
&+& y\delta_y[y\delta_y(2x+\delta_y)+
2(x-y\delta_y)c^2_w]C_0(6) \nonumber \protect\\
&+& [-x(x+2\delta_y)+2xy(x+2y)c^2_w+\delta_-\delta_+c^4_w]
[xC_0(7)+yC_0(8)]\rbrace \nonumber \protect\\
&+& \frac{m^4_Z}{4x^2y^2\delta}\lbrace [x\delta(3x(y-x\delta_x)
+2y(2x^2+y^2)-13xy^2)+x^2\delta(x(1+x^2+y^2) \nonumber \protect\\ 
&+&  2\delta_y(xy-\delta_-\delta_+))c^2_w+x(\delta_+(y(2x+7y-3)-3x\delta_x)+
8xy)c^4_w+\delta_-\delta^2_+c^6_w]D_0(1) \nonumber \protect\\
&+& [y^3\delta(x(4\delta_y-x)+\delta_y^2)+y^2\delta(3\delta_-(1+\delta_-)
-4xy)c^2_w \nonumber \protect\\
&+& y(x(x(x-7y)-2y)-3\delta_-(x+y\delta_y))c^4_w+\delta_-\delta^2_+c^6_w]
D_0(2) \nonumber \protect\\
&+& (xy-\delta_+c^2_w)[xy^2(x+2\delta_y)-2xy(x+2y)c^2_w-\delta_-\delta_+c^4_w]
D_0(3)\rbrace,
\end{eqnarray}

\begin{equation}
\label{eqaw2}
A^W_2(x,y)=-A^W_1(x\leftrightarrow y),
\end{equation}

\begin{eqnarray}
\label{eqaw3}
A^W_3(x,y) &=& \frac{1}{2xy\delta}\lbrace B_0(3)-B_0(4)+2x[B_0(4)-B_0(1)]
\rbrace \nonumber \protect\\
&+& \frac{m^2_Z}{2x^2y^2\delta^2}\lbrace -x^2\delta^2(\delta-2c^2_w)C_0(2)+
x[x\delta(x-\delta_+(x+2y))+2x(x\delta-2y)c^2_w-\delta^2_+]C_0(3) 
\nonumber \protect\\
&+& \delta_+[x^2\delta(\delta_x-5y)-2x(x\delta-2y)c^2_w+\delta_+c^4_w]
C_0(1) \nonumber \protect\\
&+& x[x\delta_x(\delta_x(\delta-y)-2y^2)-2x(x+\delta^2)c^2_w]C_0(5) 
\nonumber \protect\\
&+& x[(xy(2(x\delta_x+y\delta_y)-xy)-2xy(1+\delta)c^2_w-\delta^2_+c^4_w]
C_0(7)\rbrace \nonumber \protect\\
&+& \frac{m^4_Z}{2x^2y^2\delta^2}\lbrace -(xy-\delta_+c^2_w)[xy(\frac{1}{2}xy+
2x\delta_x)-xy(1+\delta)c^2_w-x\delta_+c^4_w]D_0(3) \nonumber \protect\\
&+& [x^2\delta^2(x\delta_x-y(3-2y))+x\delta(5xy\delta_y+y(x^2+2y^2)+3x^2
\delta_+)c^2_w \nonumber \protect\\
&+& x\delta_+(3x\delta_x-2xy+y(y-5))c^4_w-\delta^3_+c^6_w]D_0(1)\rbrace 
\nonumber \protect\\
&+& (x\leftrightarrow y), 
\end{eqnarray}

\begin{eqnarray}
\label{eqaw4}
A^W_4(x,y) &=& -\frac{2}{x\delta_+}+\frac{1}{x^2\delta_y\delta^2_+}
\lbrace 2(2x+y)[B_0(4)-B_0(3)+y(B_0(2)-B_0(4))]+x^2(B_0(2)-B_0(3))
\rbrace \nonumber \protect\\
&+& \frac{m^2}{x^3y\delta_y\delta^2}\lbrace -\delta_y[2y\delta_++
2\delta(x^2(\delta_--1)-y^2)c^2_w+y\delta_+\delta^2c^4_w]C_0(1)+ 
y\delta_y\delta^2(\delta-2c^2_w)C_0(2) \nonumber \protect\\
&+& x\delta_xc^2_w[x\delta(2y-\delta_x)+yc^2_w]C_0(3)+y\delta_x[y\delta^2+
\delta(x\delta_x+2y)c^2_w+yc^4_w]C_0(4)  \nonumber \protect\\
&+& x\delta_y\delta^3_xc^2_wC_0(5)-[y\delta\delta^2_y-(\delta^2_y(2\delta_
-\delta_++2y\delta_x-x(1+x^2))+2x^3y^2)c^2_w]C_0(6) \nonumber \protect\\    
&-& y\delta_y(x\delta_y-c^2_w)c^2_w[xC_0(7)+7C_0(8)]\rbrace 
\nonumber \protect\\
&+& \frac{m^4_Z}{x^3y\delta^2}\lbrace c^2_w[x^2\delta_x\delta^2+x\delta
(x\delta_x-y(2-x))c^2_w+y\delta_+c^2_w]D_0(1) \nonumber \protect\\
&-& [\delta^2_y\delta^3-y\delta^2(x(x-2)+3y)c^2_w+\delta(x\delta_x\delta_-+
3y^2)c^4_w-y\delta_+c^6_w]D_0(2) \nonumber \protect\\
&+& y(x\delta_x-c^2_w)(xy-\delta_+c^2_w)c^2_wD_0(3)\rbrace,
\end{eqnarray}

\begin{eqnarray}
\label{eqaw5}
A^W_5(x,y) &=& \frac{2}{x\delta_+}+\frac{1}{x^2y\delta\delta^2_+}\lbrace 
[\delta_+(y(3x-2\delta_y)-x^2)-2xy][B_0(3)-B_0(4)] \nonumber \protect\\
&+& \delta^2_+[y(x-2\delta_y)(B_0(4)-B_0(2))+x^2(B_0(1)-B_0(4))]
\rbrace \nonumber \protect\\
&+& \frac{m^2_Z}{2x^3y^2\delta_+\delta^2}\lbrace \delta_+\delta^2
[\delta(x(x^2+5y^2)+2y^2\delta)-2(x(x^2+y^2)+2y^2\delta)c^2_w
\nonumber \protect\\ &-& 2\delta^2_+
(2\delta y^2+x\delta_-\delta_+)c^4_w]C_0(1) \nonumber \protect\\
&+& [\delta^2_+\delta^2(x(x^2+5y^2)+2y^2\delta)-2(x\delta_+\delta_-+
2y^2\delta)c^4_w+2\delta_+(\delta\delta_+(x^2+3y^2) \nonumber \protect\\
&+& 2y(\delta^2(x^2+y^2)-2x^2\delta_+))c^4_w]C_0(2) \nonumber \protect\\
&+& x\delta_+[x^3\delta^2+2x(-x\delta\delta_-+2xy\delta_+-y\delta)c^2_w+
(x\delta_-\delta_++2y^2\delta)c^4_w]C_0(3) \nonumber \protect\\
&+& y\delta_+[y^2\delta^2(5x+2\delta)+2y(2x^2\delta_++\delta(x-2\delta_x))
c^2_w+(x\delta_-\delta_++2y^2\delta)c^4_w]C_0(4) \nonumber \protect\\
&+& x\delta_+[x^2\delta_x(3y^2-\delta^2)+2(x^2(y\delta_x+\delta^2)-
y\delta(2x-1))c^2_w]C_0(5) \nonumber \protect\\
&+& y\delta_+\delta_y[y(x(3x^3-5\delta^2)-2\delta^3)+2(x^3-\delta_-
\delta^2)c^2_w]C_0(6) \nonumber \protect\\
&+& [-3x^3y^2+2xy(2y(x-\delta)+x^2)c^2_w+(x\delta_-\delta_++2y^2\delta)
c^4_w][xC_0(7)+yC_0(8)]\rbrace \nonumber \protect\\
&+& \frac{M^4_Z}{2x^3y^2\delta^2}\lbrace [-x^4\delta^3+x^2\delta(\delta(
3x\delta_-+2y(1-2y)) -4xy\delta_-)c^2_w \nonumber \protect\\
&+& x(x\delta^2_+(4y-3\delta)+3\delta_-\delta_+-x\delta+2y\delta_x)c^4_w+
\delta_+(x\delta_-\delta_++2y^2\delta)c^6_w]D_0(1) \nonumber \protect\\
&+& [-y^3\delta^3(3x+2y\delta)+y^2\delta(4x^2\delta_-\delta(x(5x-3y+4)-
6y))c^2_w+(-3xy\delta(2x+y) \nonumber \protect\\
&+& 2\delta^2(3y^2-x\delta_-)+x^2(4(x^2+y^2)+x(7y-\delta_x)))c^4_w]D_0(2) 
\nonumber \protect\\
&+& (xy-c^2_w\delta_+)[(3x^3y^2-2xy(2y(2x-\delta_y)+x^2))c^2_w
\nonumber \protect\\
&+&(x(3y^2-x^2)-2y^2\delta_y)c^4_w]D_0(3)\rbrace,                                        
\end{eqnarray}

\begin{equation}
\label{eqaw6}
A^W_6(x,y)=-A^W_4(x\leftrightarrow y),
\end{equation}

\begin{equation}
\label{eqaw7}
A^W_7(x,y)=-A^W_5(x\leftrightarrow y),
\end{equation}

\noindent with $B_0(i)$, $C_0(i)$, and $D_0(i)$ the two--, three-- and
four--point scalar functions given in the Appendix. 

\section{Numerical results and discussion}
\label{num}

\begin{figure}
\centerline{ \epsfysize=8cm \epsfbox{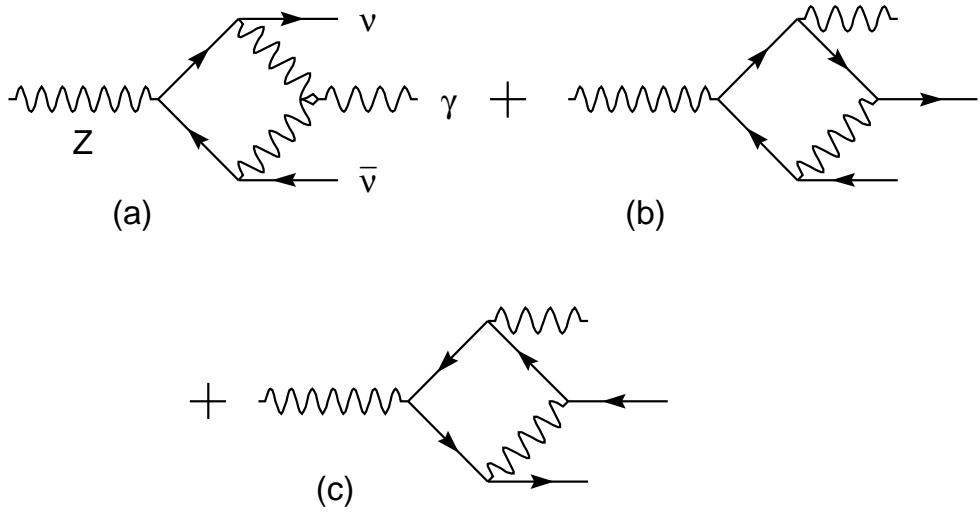} }
\caption{Box diagrams with $Zl^+l^-$ coupling.}
\label{fig4}
\end{figure}

\begin{figure}
\centerline{ \epsfysize=8cm \epsfbox{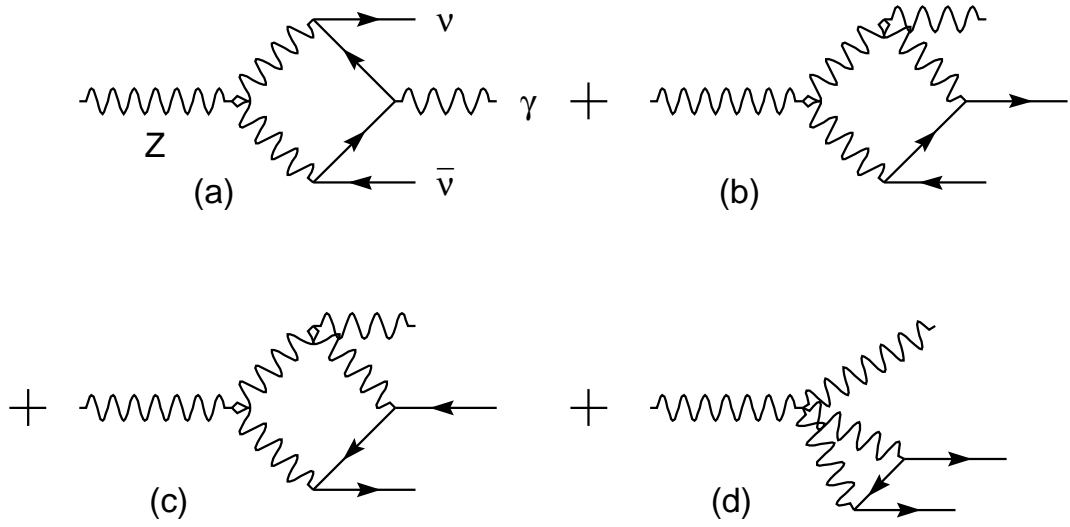} }
\caption{Box diagrams with $ZW^+W^-$ coupling.}
\label{fig5}
\end{figure}

In the previous section the amplitudes for the sets of diagrams shown in
Figs. \ref{fig2}--\ref{fig5} have been arranged in an $\mathrm{U_e(1)}$
gauge invariant form. We have shown that the respective amplitudes are
$\mathrm{SU(2)}$ gauge independent and $\mathrm{U_e(1)}$ gauge invariant
for each set of diagrams in Figs. \ref{fig2} and \ref{fig3}.  This
property enables us to discuss separately each contribution. In
particular, we would like to study the role played by the off--shell
$\mathrm{U_e(1)}$ gauge structures of the $Z$ boson and the neutrino. From
(\ref{eqwidth}) and (\ref{eqamp}) the double phase space integral for the
decay width can be expressed as

\begin{eqnarray}
\label{eqw}
\Gamma(Z \to \bar{\nu}\nu\gamma) &=& 
\frac{m_Z}{256\pi^3}\int_{0}^{1}dx\int_{0}^{1-x}dy \lbrace 
\arrowvert{\cal M}_{Z^*}\arrowvert^2+\arrowvert{\cal M}_{\nu^*}
\arrowvert^2+\arrowvert{\cal M}_{Box4}\arrowvert^2
+\arrowvert {\cal M}_{Box5}\arrowvert^2 \nonumber \protect\\
&+& 2Re[({\cal M}_{\nu^*}+{\cal M}_{Box4}+{\cal M}_{Box5}){\cal M}^{\dag}_
{Z^*}+({\cal M}_{Box4}+ {\cal M}_{Box5}){\cal M}^{\dag}_{\nu^*}
\nonumber \protect\\
&+& {\cal M}_{Box4}{\cal M}^{\dag}_{Box5}]\rbrace.
\end{eqnarray}

In order to disentangle the relative importance of the contributions
arising from the $ZZ^*\gamma$ and $\nu\nu^*\gamma$ couplings as well as
the box diagrams, we have evaluated separetely the corresponding terms in
(\ref{eqw}). In the case of the box diagrams contributions, we have
obtained separetely the contribution arising from each of the five
$\mathrm{U_e(1)}$ gauge structures defining these amplitudes. The result
for the branching ratio $B(Z \to \bar{\nu}\nu \gamma)$ will follow
inmediately.\protect\\

A calculation of this type requires a careful numerical evaluation of the
Passarino--Veltman scalar functions, in particular of those arising from
diagrams with resonant effects, as it is the case for the graphs shown in
Fig.  \ref{fig4}. Since some of the involved scalar functions can not be
expressed in terms of elementary functions, it is necessary to evaluate
them by means of more elaborated methods, such as the Fortran program FF
\cite{FF}. We have made this task in different ways to get a reliable
answer. In the case of the $ZZ^*\gamma$ and $\nu\nu^*\gamma$ couplings
contributions, we get simple analytical solutions for the corresponding
amplitudes, so we programmed the required integrals in Mathematica
\cite{wol} and Fortran language together with the FF package. We have
found an excellent agreement between the obtained results. In the case of
the box diagrams, the four--point functions were analized by several
methods. Explicit solutions of this functions were numerically evaluated
and the results were compared with the corresponding evaluation using the
FF package. In this way we are certain \cite{Olden} that we get the
correct answer for these functions. We found delicate cancellations
arising from the box amplitudes and in order to get the required accuracy
different integration methods were used.\protect\\

Not all the box diagrams present troubles for their numerical evaluation. 
The most serious difficulties arise from the box diagrams shown in Fig. 
\ref{fig4}. In particular, the graphs (b) and (c)  of this set induce
collinear singularities. This fact implies, on one hand, that we must keep
the lepton masses within the arguments of the scalar functions and, on the
other hand, that we should have special care with the numerical evaluation
of the corresponding amplitude. In order to get a reliable answer for
these diagrams, we have used an entirely independent check. The idea is
just to evaluate the amplitude via unitarity conditions

\begin{equation} 
\label{equ} 
Im {\cal M}_{Box4}=\frac{1}{2}
\int d\Phi_l {\cal M}(Z \to l^+l^-) 
{\cal M}^{\dag}(\bar{\nu}\nu\gamma \to l^+l^-), 
\end{equation}

\noindent where $d\Phi_l$ is the phase space for the intermediate lepton
pair. From the tree--level amplitudes of the right--hand side of this
equation we can see that only the graphs \ref{fig4}(b) and \ref{fig4}(c)
have collinear sigularities. The result for the left--hand side of
(\ref{equ}), obtained by the above described method, agrees nicely with
the one obtained for the right--hand side.\protect\\
 
We now proceed to discuss our results \footnote{The values used for the SM
parameters were taken from the Particle Data Group \cite{PDG}.}. First,
the $\nu\nu^*\gamma$ coupling contribution to the decay width $ \Gamma({Z
\to \bar{\nu}\nu \gamma})$ is $2.02 \times 10^{-10}$ GeV. On the other
hand, in the case of the $ZZ^* \gamma$ coupling only the top quark family
gives an important contribution, namely $5.74 \times 10^{-12}$ GeV, while
the remaining contributions are negligible small.  This fact can be
understood because the $ZZ^*\gamma$ coupling contributions have a strong
dependence on the mass gap existing in each fermion family.  In
particular, an exact mass degeneration implies a zero contribution. This
nondecoupling effect of chiral fermions is well known in the SM and it is
due to isospin breaking.  This constitutes an important effect in
radiative corrections proportional to $m_t^2-m_b^2$.  In the same footing,
it is interesting to study a possible enhancement for the $ZZ^*\gamma$
amplitude arising from a fourth family of chiral fermions with SM
assignment of quantum numbers.  The most promising scenario arises from a
heavy top--like quark with a light bottom--like quark and a light
tau--like lepton. In order to give an estimation of this contribution, we
have used the LEP1 lower bounds for the tau--like lepton and the
bottom--like quark:  $m'_{\tau}\le 80$ GeV and $m'_b \le 46$ GeV; also, a
variation of $m'_t$ in the 300--850 GeV range is assumed, where the upper
limit for the top--like quark mass was taken from the $\Delta \rho$
constraint over the fourth family mass difference \cite{Gun94}. The
corresponding contribution, as a function of $\Delta m=m'_t-m'_b$ shows an
enhancement of almost 5 times with respect to the third family
contribution, this enhancement is rapidly reached at $\Delta m=300$ GeV.
Finally, in the case of the box diagrams as well as the interference
terms, their contribution is of the same order as that coming from the
$\nu\nu^*\gamma$ coupling. We have found that the main contribution of the
box diagrams arises from the $\mathrm{U_e(1)}$ gauge structure
corresponding to the neutrino MDT, whereas the $\mathrm{U_e(1)}$ gauge
structures given in (\ref{eqgs})  lead to a contribution of order
$10^{-11}$--$10^{-13}$ GeV. \protect\\

In summary, considering all the terms in (\ref{eqw}) and assuming lepton
universality we have $B(Z \to\bar{\nu}\nu \gamma)=7.16 \times 10^{-10}$
for the three SM neutrino species.  The main contribution comes from the
neutrino MDT $\mathrm{U_e(1)}$ gauge structure, while the $Z$ boson EDT
contribution is two orders of magnitude below.\protect\\
 
\section{Final remarks} 
\label{sum} 

We have presented a complete calculation of the SM decay $Z \to
\bar{\nu}\nu \gamma$ using a nonlinear gauge. We have found that this
decay is characterized by the off--shell electromagnetic properties of the
neutrino and the $Z$ boson: a neutrino MDT and a $Z$ boson EDT induced by
the $\nu\nu^*\gamma$ and the $ZZ^*\gamma$ couplings, respectively. We
found that the $Z$ boson EDT contribution may be enhanced by a fourth
generation of chiral fermions; however, this effect is not enough to
increase significantly $B(Z \to \bar{\nu}\nu \gamma$). On the other hand,
the main contribution to the decay $Z \to \bar{\nu}\nu \gamma$ arises from
the neutrino MDT. This contribution depends strongly on the momentum
transferred by the $Z$ boson but it is not sensitive to a possible
neutrino mass. Another interesting fact found in our calculation is that
the box diagrams contributions can not be neglected because they induce a
neutrino MDT $\mathrm{U_e(1)}$ gauge structure with a form factor of the
same order of magnitude as that coming from the $\nu\nu^*\gamma$ coupling. 
It is important to notice that this is true even though there is no
$\mathrm{SU(2)}$ nor $\mathrm{U_e(1)}$ gauge link between the
$\nu\nu^*\gamma$ and box diagrams. This scenario was foreseen in a recent
study of the decay $Z \to \bar{\nu}\nu \gamma$ within the effective
Lagrangian approach \cite{Perez}: the operators inducing the $ZZ^*\gamma$
coupling are suppressed with respect to the dimension--six and
dimension--eight operators inducing the $\nu\nu^*\gamma$ coupling and the
box diagram contributions. Therefore, it is expected that the
dimension--six and dimension--eight operators contribute substantially to
$B(Z \to \bar{\nu}\nu \gamma$), as it was the case in our calculation. 
Finally, we would like to point out that the decay $Z \to \bar{\nu}\nu
\gamma$ leads to a branching ratio of the same order of magnitude than the
one expected for the radiative decay $Z \to \gamma\gamma\gamma$
\cite{z3p}. In this case, the experimental bound for the respective
branching ratio is also about four orders of magnitude below the SM
prediction. Therefore, these decay modes of the $Z$ boson should be of
interest for people searching for deviations coming from new physics
effects.\protect\\

\acknowledgments {We appreciate useful discussions with J. L. Diaz-Cruz
and F. Larios. MAP acknowledges the hospitality of the Aspen Center for
Physics, where part of this work was done. Support from CONACYT (M\'exico)
is aknowledged.}

\section*{Appendix:  Passarino--Veltman scalar functions} 
\label{apI} 
In this appendix we present the arguments of the scalar functions involved
in the $Z\bar{\nu}\nu\gamma$ interaction. We follow the notation of
\cite{Mer91} and the lepton mass was denoted by $m_l$. \protect\\

\setcounter{equation}{0}

Two--point scalar functions
\begin{eqnarray}
B_0(1) &=& B_0(xm^2_Z,m^2_l,m^2_W), \nonumber \protect\\
B_0(2) &=& B_0(ym^2_Z,m^2_l,m^2_W), \nonumber \protect\\
B_0(3) &=& B_0(m^2_Z,m^2_W,m^2_W), \nonumber \protect\\
B_0(4) &=& B_0(\delta m^2_Z,m^2_W,m^2_W), \nonumber \protect\\
B_0(5) &=& B_0(m^2_Z,m^2_l,m^2_l), \nonumber \protect\\
B_0(6) &=& B_0(\delta m^2_Z,m^2_l,m^2_l). 
\end{eqnarray}\protect\\

Three--point scalar functions
\begin{eqnarray}
C_0(1) &=& C_0(0,\delta m^2_Z,m^2_Z,m^2_W,m^2_W,m^2_W), \nonumber \protect\\
C_0(2) &=& C_0(0,0,\delta m^2_Z,m^2_W,m^2_l,m^2_W), \nonumber \protect\\
C_0(3) &=& C_0(0,0,xm^2_Z,m^2_W,m^2_W,m^2_l), \nonumber \protect\\
C_0(4) &=& C_0(0,0,ym^2_Z,m^2_W,m^2_W,m^2_l), \nonumber \protect\\
C_0(5) &=& C_0(0,xm^2_Z,m^2_Z,m^2_W,m^2_l,m^2_W), \nonumber \protect\\
C_0(6) &=& C_0(0,ym^2_Z,m^2_Z,m^2_W,m^2_l,m^2_W), \nonumber \protect\\
C_0(7) &=& C_0(0,0,xm^2_Z,m^2_W,m^2_l,m^2_l), \nonumber \protect\\
C_0(8) &=& C_0(0,0,ym^2_Z,m^2_W,m^2_l,m^2_l), \nonumber \protect\\
C_0(9) &=& C_0(0,xm^2_Z,m^2_Z,m^2_l,m^2_W,m^2_l), \nonumber \protect\\
C_0(10) &=& C_0(0,ym^2_Z,m^2_Z,m^2_l,m^2_W,m^2_l), \nonumber \protect\\
C_0(11) &=& C_0(0,0,\delta m^2_Z,m^2_l,m^2_W,m^2_l).
\end{eqnarray}\protect\\

Four--point scalar functions
\begin{eqnarray}
D_0(1) &=& D_0(0,0,0,m^2_Z,xm^2_Z,\delta m^2_Z,m^2_W,m^2_W,m^2_l,m^2_W),
 \nonumber \protect\\
D_0(2) &=& D_0(0,0,0,m^2_Z,ym^2_Z,\delta m^2_Z,m^2_W,m^2_W,m^2_l,m^2_W),
 \nonumber \protect\\
D_0(3) &=& D_0(0,0,0,m^2_Z,xm^2_Z,ym^2_Z,m^2_W,m^2_l,m^2_l,m^2_W), 
\nonumber \protect\\
D_0(4) &=& D_0(0,0,0,m^2_Z,xm^2_Z,ym^2_Z,m^2_l,m^2_W,m^2_W,m^2_l), 
\nonumber \protect\\
D_0(5) &=& D_0(0,0,0,m^2_Z,xm^2_Z,\delta m^2_Z,m^2_l,m^2_l,m^2_W,m^2_l), 
\nonumber \protect\\
D_0(6) &=& D_0(0,0,0,m^2_Z,m^2_Z,\delta m^2_Z,m^2_l,m^2_W,m^2_l,m^2_l).
\end{eqnarray}

It is important to notice that lepton mass can be droped out
from the above equations, except in $C_0(7)$, $D_0(5)$ and $D_0(6)$, where
collinear singularities arise.\protect\\

\end{document}